\documentclass{article}
\usepackage{graphicx}
\usepackage{amsmath}

\begin{document}

\noindent \textbf{Lorentz Invariant\ Majorana Formulation of the Field
Equations and}

\noindent \textbf{Dirac-like Equation for the Free Photon }\bigskip \bigskip

\qquad Tomislav Ivezi\'{c}

\qquad\textit{Ru%
\mbox
{\it{d}\hspace{-.15em}\rule[1.25ex]{.2em}{.04ex}\hspace{-.05em}}er Bo\v
{s}kovi\'{c} Institute, P.O.B. 180, 10002 Zagreb, Croatia}

\textit{\qquad ivezic@irb.hr}\bigskip \bigskip

\noindent In this paper we present a new geometric formulation (Clifford
algebra formalism) of the field equations, which is independent of the
reference frame and of the chosen system of coordinates in it. This
formulation deals with the complex 1-vector $\Psi =E-icB$ ($i$ is the unit
imaginary), which is four-dimensional (4D) geometric generalization of
Majorana's complex 3D quantity $\mathbf{\Psi }=\mathbf{E}-ic\mathbf{B}$.
When the sources are absent the field equations with the complex $\Psi $
become Dirac-like relativistic wave equations for the free photon. In the
frame of ``fiducial'' observers (the observers who measure fields are at
rest) and in the standard basis the component form of the field equations
with 4D $\Psi $ reproduces the component form of Majorana-Maxwell equations
with 3D field $\mathbf{\Psi }$. The important differences between the
approach with the 4D $\Psi $ and that one with the 3D $\mathbf{\Psi }$ are
discussed. \medskip \bigskip

\noindent Keywords: \textit{Lorentz invariant\ field equations -
Majorana-Maxwell equations\bigskip }

\noindent PACS: 03.30.+p, 03.50.De\bigskip \medskip

\noindent \textbf{1. Introduction}\bigskip

\noindent In Majorana formulation of electrodynamics the Maxwell equations
are written in terms of complex combination of the three-dimensional (3D)
vectors of the electric and the magnetic fields $\mathbf{E}$ and $\mathbf{B}$
respectively, $\mathbf{\Psi }=\mathbf{E}-ic\mathbf{B}$, see [1,2]. (The
vectors in the 3D space will be designated in bold-face.) In terms of $%
\mathbf{\Psi }$ the Maxwell equations in vacuum can be cast in a Dirac-like
form using the correspondence principle $W\rightarrow i$%
h\hskip-.2em\llap{\protect\rule[1.1ex]{.325em}{.1ex}}\hskip.2em%
$\partial /\partial t$, $\mathbf{p}\rightarrow -i$%
h\hskip-.2em\llap{\protect\rule[1.1ex]{.325em}{.1ex}}\hskip.2em%
$\mathbf{\nabla }$. In that case $\mathbf{\Psi }^{\ast }\mathbf{\cdot \Psi =E%
}^{2}+c^{2}\mathbf{B}^{2}$ is proportional to the probability density
function for a photon. An important advantage of Majorana formulation of
electrodynamics is that it does not make use of the intermediate
electromagnetic potentials but deals with observable quantities, the
electric and the magnetic fields.

Covariant Majorana formulation is developed in [2]. There the covariant form
of the complex field $\Psi _{\mu }=E_{\mu }-iB_{\mu }$ is introduced. The
covariant Maxwell equation with $\Psi _{\mu }$ are written only for the free
fields, i.e., when $j^{\beta }=0.$ It is worth noting that $E_{\mu }$, $%
B_{\mu }$ and $\Psi _{\mu }$ are components that are determined in the
specific system of coordinates, which we call Einstein's system of
coordinates. In Einstein's system of coordinates the standard, i.e.,
Einstein's synchronization [3] of distant clocks and Cartesian space
coordinates $x^{i}$ are used in the chosen inertial frame. We also point out
that in [2] $E^{\mu }$ and $B^{\mu }$ are treated as the ''auxiliary
fields,'' while the 3D vectors $\mathbf{E}$ and $\mathbf{B}$ are considered
as the physical fields.

Further generalization of Majorana formulation is presented in [4]. There a
geometric approach to special relativity is developed, which deals with
tensors as 4D geometric quantities. We note that such geometric approach
with tensors as geometric quantities is considered not only in [4] but in
[5, 6] as well, while a similar treatment in which 4D geometric quantities
are Clifford multivectors is presented in [7-10]. The approach to special
relativity with 4D geometric quantities is called the invariant special
relativity. In the the invariant special relativity one considers that the
4D geometric quantities are well-defined both theoretically and \emph{%
experimentally} in the 4D spacetime, and not, as usual, the 3D quantities.
All physical quantities are defined without reference frames, i.e., as
absolute quantities (AQs) or, when some basis has been introduced, they are
represented as 4D coordinate-based geometric quantities (CBGQs) comprising
both components and a \emph{basis}. It is shown in the mentioned references
that such geometric approach is in a complete agreement with the principle
of relativity and, what is the most important, with experiments, see [5]
(tensor formalism) and [8-10] (geometric algebra formalism). In [4] Sec. 6.3
the invariant Majorana electromagnetic field $\Psi ^{a}$ is defined as $\Psi
^{a}=E^{a}-icB^{a}$, where $E^{a},$ $B^{a}$ and $\Psi ^{a}$ are the 4D AQs
with definite physical meaning and not the ''auxiliary fields''. In the same
section the field equation with $\Psi ^{a}$ is presented, which for $%
j^{\beta }=0$ is reduced to the Dirac-like relativistic wave equation for
the free photon.

In this paper we shall explore a similar Lorentz invariant Majorana
formulation in which physical quantities will be represented by Clifford
multivectors. To simplify the derivation of all important relations we shall
employ recently developed axiomatic geometric formulation of
electromagnetism [10] in which the primary quantitity for the whole
electromagnetism is the electromagnetic field $F$ (bivector). New Lorentz
invariant Majorana form of the field equations and Dirac-like equations for
the free photon are reported. The similarities and the differences between
our Lorentz invariant field equations with the 4D $\Psi $ and
Majorana-Maxwell equations with the 3D\textbf{\ }$\mathbf{\Psi }$ are
discussed.\bigskip

\noindent \textbf{2. A brief summary of geometric algebra }\bigskip

\noindent In this paper the investigation with 4D geometric quantities will
be done in the geometric algebra formalism, see, e.g., [11] and [12]. First
we provide a brief summary of Clifford algebra with multivectors. Clifford
vectors are written in lower case ($a$) and general multivectors (Clifford
aggregate) in upper case ($A$). The space of multivectors is graded and
multivectors containing elements of a single grade, $r$, are termed
homogeneous and often written $A_{r}.$ The geometric (Clifford) product is
written by simply juxtaposing multivectors $AB$. A basic operation on
multivectors is the degree projection $\left\langle A\right\rangle _{r}$
which selects from the multivector $A$ its $r-$ vector part ($0=$ scalar, $%
1= $ vector, $2=$ bivector, ....). We write the scalar (grade-$0$) part
simply as $\left\langle A\right\rangle .$ The geometric product of a grade-$%
r $ multivector $A_{r}$ with a grade-$s$ multivector $B_{s}$ decomposes into
$A_{r}B_{s}=\left\langle AB\right\rangle _{\ r+s}+\left\langle
AB\right\rangle _{\ r+s-2}...+\left\langle AB\right\rangle _{\ \left|
r-s\right| }.$ The inner and outer (or exterior) products are the
lowest-grade and the highest-grade terms respectively of the above series $%
A_{r}\cdot B_{s}\equiv \left\langle AB\right\rangle _{\ \left| r-s\right| },$
and $A_{r}\wedge B_{s}\equiv \left\langle AB\right\rangle _{\ r+s}.$ For
vectors $a$ and $b$ we have $ab=a\cdot b+a\wedge b,$ where $a\cdot b\equiv
(1/2)(ab+ba),$ and $a\wedge b\equiv (1/2)(ab-ba).$ In the case considered in
this paper Clifford algebra is developed over the field of the complex
numbers. Complex reversion is the operation which takes the complex
conjugate of the scalar (complex) coefficient of each of the 16 elements in
the algebra and reverses the order of multiplication of vectors in each
multivector, $\overline{AB}=\overline{B}\overline{A}$, where, e.g., the
complex reversion of $A$ is denoted by an overbar $\overline{A}$.

Any multivector $A$ is a geometric 4D quantity defined without reference
frame, i.e., an AQ. When some basis has been introduced $A$ can be written
as a CBGQ comprising both components and a basis. Usually [11, 12] one
introduces the standard basis. The generators of the spacetime algebra are
taken to be four basis vectors $\left\{ \gamma _{\mu }\right\} ,\mu =0,...3$
(the standard basis) satisfying $\gamma _{\mu }\cdot \gamma _{\nu }=\eta
_{\mu \nu }=diag(+---).$ This basis is a right-handed orthonormal frame of
vectors in the Minkowski spacetime $M^{4}$ with $\gamma _{0}$ in the forward
light cone. The $\gamma _{k}$ ($k=1,2,3$) are spacelike vectors. The basis
vectors $\gamma _{\mu }$ generate by multiplication a complete basis for the
spacetime algebra: $1,\gamma _{\mu },\gamma _{\mu }\wedge \gamma _{\nu
},\gamma _{\mu }\gamma _{5,}\gamma _{5}$ ($16$ independent elements). $%
\gamma _{5}$ is the pseudoscalar for the frame $\left\{ \gamma _{\mu
}\right\} .$

Observe that the standard basis corresponds, in fact, to Einstein's system
of coordinates. However different systems of coordinates are allowed in an
inertial frame and they are all equivalent in the description of physical
phenomena. For example, in [4], two very different, but physically
completely equivalent, systems of coordinates, Einstein's system of
coordinates and the system of coordinates with a nonstandard
synchronization, the everyday (radio) (``r'') synchronization, are exposed
and exploited throughout the paper. In order to treat different systems of
coordinates on an equal footing we have developed such form of the LT which
is independent of the chosen system of coordinates, including different
synchronizations, [4] (tensor formalism) and [7] (Clifford algebra
formalism). Furthermore in [4] we have presented the transformation matrix
that connects Einstein's system of coordinates with another system of
coordinates in the same reference frame. For the sake of brevity and of
clearness of the whole exposition, we shall write CBGQs only in the standard
basis $\left\{ \gamma _{\mu }\right\} $, but remembering that the approach
with 4D geometric quantities holds for any choice of basis. All equations
written with 4D AQs and 4D CBGQs will be manifestly Lorentz invariant
equations.

As mentioned above a Clifford multivector $A$, an AQ, can be written as a
CBGQ, thus with components and a basis. Any CBGQ is an invariant quantity
under the passive Lorentz transformations; both the components and the basis
vectors are transformed but the whole 4D geometric quantity remains
unchanged, e.g., the position 1-vector $x$ can be decomposed in the $S$ and $%
S^{\prime }$ (relatively moving) frames and in the standard basis $\left\{
\gamma _{\mu }\right\} $ and some non-standard basis $\left\{ r_{\mu
}\right\} $ as $x=x^{\mu }\gamma _{\mu }=x^{\prime \mu }\gamma _{\mu
}^{\prime }=....=x_{r}^{\prime \mu }r_{\mu }^{\prime }.$ The primed
quantities are the Lorentz transforms of the unprimed ones. The invariance
of some 4D CBGQ under the passive Lorentz transformations reflects the fact
that such mathematical, invariant, geometric 4D quantity represents the same
physical quantity for relatively moving observers. \bigskip \medskip

\noindent \textbf{3.}\textit{\ }\textbf{The relations that connect }$F$
\textbf{with} $E$, $B$ \textbf{and with}\textit{\ }$\Psi $, $\overline{\Psi }
$ \bigskip

\noindent In contrast to the usual Clifford algebra approaches [11, 12], and
all other previous approaches, we have shown in [10] that the bivector field
$F$, and not the 3D vectors $\mathbf{E}$ and $\mathbf{B}$ or the
electromagnetic potentials, can be considered as the primary physical
quantity for the whole electromagnetism. From the known $F$ one can find
different 4D quantities that represent the 4D electric and magnetic fields;
they are considered in [8] and [9]. One of these representations, which is
examined in [7-9], uses the decomposition of $F$ into 1-vectors $E$ and $B$

\begin{align}
F& =(1/c)E\wedge v+(B\wedge v)I,  \notag \\
E& =(1/c)F\cdot v,\quad B=(1/c^{2})(F\wedge v)I;\ E\cdot v=B\cdot v=0,
\label{itf}
\end{align}
where $I$ is the unit pseudoscalar. ($I$ is defined algebraically without
introducing any reference frame, as in [13] Sec. 1.2.) The velocity $v$ can
be interpreted as the velocity (1-vector) of a family of observers who
measures $E$ and $B$ fields. That velocity $v$ and all other quantities
entering into (\ref{itf}) are defined without reference frames, i.e., they
are AQs.

It is proved in [8, 9] (Clifford algebra formalism) and [6] (tensor
formalism) that the observers in relative motion see the same field, e.g.,
the $E$ field in the $S$ frame is the same as in the relatively moving $%
S^{\prime }$ frame; $E^{\mu }\gamma _{\mu }=E^{\prime \mu }\gamma _{\mu
}^{\prime }$, where all primed quantities are the Lorentz transforms of the
unprimed ones. The LT transform the components $E^{\mu }$ from the $S$ frame
again to the components $E^{\prime \mu }$ from the $S^{\prime }$ frame, in
the same way as for any other 1-vector. For example, the transformations for
the components of the 1-vector $E$ are
\begin{equation}
E^{\prime 0}=\gamma (E^{0}-\beta E^{1}),E^{\prime 1}=\gamma (E^{1}-\beta
E^{0}),E^{\prime 2}=E^{2},E^{\prime 3}=E^{3},  \label{ea}
\end{equation}
and the same for the transformations of the components of the 1-vector $B$.
Thus the Lorentz transformed $E^{\prime \mu }$ are not expressed by the
mixture of components $E^{\mu }$ and $B^{\mu }$ of the electric and magnetic
fields respectively from the $S$ frame. This is in sharp contrast to all
previous formulations of electromagnetism, starting with Einstein's work
[3], in which the components $E_{i}^{\prime }$ of the 3D $\mathbf{E}^{\prime
}$ are expressed by the mixture of components of $E_{i}$ and $B_{i}$ from
the $S$ frame. For example, the transformations for the components of the 3D
$\mathbf{E}$ are
\begin{equation}
E_{x}^{\prime }=E_{x},\ E_{y}^{\prime }=\gamma (E_{y}-\beta cB_{z}),\
E_{z}^{\prime }=\gamma (E_{z}+\beta cB_{y}),  \label{et}
\end{equation}
and similarly for the components of the 3D $\mathbf{B}$, see, e.g., [14]
Sec. 11.10. In all textbooks and papers treating relativistic
electrodynamics these usual transformations of the components of the 3D $%
\mathbf{E}$ (\ref{et}) and $\mathbf{B}$ (e.g., [14] Sec. 11.10) are
considered to be the LT, but the fundamental results obtained in [6] and [8,
9] exactly show that they drastically differ from the LT of the 4D
quantities that represent the electric and the magnetic fields.

Next we introduce the complex fields, the 4D AQ $\Psi $ and its complex
reversion $\overline{\Psi }$. They are defined in terms of 1-vectors of the
electric and magnetic fields $E$ and $B$ as
\begin{align}
\Psi & =E-icB,\quad \overline{\Psi }=E+icB,  \notag \\
E& =(1/2)(\Psi +\overline{\Psi }),\quad B=(i/2c)(\Psi -\overline{\Psi });\
v\cdot \Psi =v\cdot \overline{\Psi }=0.  \label{pkom}
\end{align}
The complex $\Psi $ and $\overline{\Psi }$ are homogeneous, grade-1,
multivectors. The meanings of $v$ and $I$ are the same as in (\ref{itf}).

Using (\ref{itf}) we find that the $F$ formulation and the complex $\Psi $
formulation are connected by the relations
\begin{align}
F& =(1/2c)\{(\Psi +\overline{\Psi })\wedge v+i[(\Psi -\overline{\Psi }%
)\wedge v]I\}.  \notag \\
\Psi & =(1/c)F\cdot v+(i/c)I(F\wedge v).  \label{f}
\end{align}
We note that one can construct the formulation of electrodynamics with the
complex 1-vectors $\Psi $ and $\overline{\Psi }$ as 4D AQs, i.e., Lorentz
invariant Majorana formulation of electrodynamics using the relations (\ref
{f}) and the work [10]. Such formulation is perfectly suited for the
transition to the quantum electrodynamics.\bigskip \medskip

\noindent \textbf{4}\textit{. }\textbf{Lorentz invariant Majorana form of
the field equation and }

\textbf{Dirac-like equation for the free photon}\bigskip

\noindent As already mentioned we shall use the $F$ formulation [10] to find
the field equation for $\Psi $. In the $F$ formulation [10] the
electromagnetic field is represented by a bivector-valued function $F=F(x)$
on the spacetime. The source of the field is the electromagnetic current $j$
which is a 1-vector field and the gradient operator $\partial $ is also
1-vector. A single field equation for $F$ was first given by M. Riesz [15]
as
\begin{equation}
\partial F=j/\varepsilon _{0}c,\quad \partial \cdot F+\partial \wedge
F=j/\varepsilon _{0}c.  \label{MEF}
\end{equation}
The trivector part is identically zero in the absence of magnetic charge.

Using (\ref{f}) and (\ref{MEF}) we write the field equation in terms of the
complex 1-vector $\Psi $ as
\begin{equation}
\partial \cdot (\Psi \wedge v)+i\left[ \partial \wedge (\Psi \wedge v)\right]
I=j/\varepsilon _{0}.  \label{mecp}
\end{equation}
This form of the field equation (in which $\overline{\Psi }$ does not
appear) is achieved separating vector and trivector parts and then combining
them to eliminate $\overline{\Psi }$. The equation (\ref{mecp}) is the most
general basic equation for the Lorentz invariant Majorana formulation of
electrodynamics.

From this field equation with AQs one can get more familiar field equation
with CBGQs that are written in the standard basis $\left\{ \gamma _{\mu
}\right\} $. Thus instead of (\ref{mecp}) we have
\begin{equation}
\partial _{\alpha }[(\delta _{\quad \mu \nu }^{\alpha \beta }-i\varepsilon
_{\quad \mu \nu }^{\alpha \beta })\Psi ^{\mu }v^{\nu }]\gamma _{\beta
}=(j^{\beta }/\varepsilon _{0})\gamma _{\beta },  \label{meko}
\end{equation}
where $\delta _{\quad \mu \nu }^{\alpha \beta }=\delta _{\,\,\mu }^{\alpha
}\delta _{\,\,\nu }^{\beta }-\delta _{\,\,\nu }^{\alpha }\delta _{\,\mu
}^{\beta }$. The equation (\ref{meko}) can be also written as
\begin{align}
\partial _{\alpha }[(\Gamma ^{\alpha })_{\ \mu }^{\beta }\Psi ^{\mu }]\gamma
_{\beta }& =(j^{\beta }/\varepsilon _{0})\gamma _{\beta },  \notag \\
(\Gamma ^{\alpha })_{\ \mu }^{\beta }& =\delta _{\quad \nu \rho }^{\alpha
\beta }v^{\rho }g_{\ \mu }^{\nu }+i\varepsilon _{\quad \nu \mu }^{\alpha
\beta }v^{\nu }.  \label{gako}
\end{align}
We note that the same equation as (\ref{gako}) is obtained in the tensor
formalism in [4]. Observe that our $(\Gamma ^{\alpha })_{\ \mu }^{\beta }$
differ from the expression for the corresponding quantity $(\gamma ^{\alpha
})_{\ \mu }^{\beta }$, Eq. (30) in [2].

In the case when the sources are absent, $j=0$, and when it is assumed that
the velocity 1-vector $v$ is independent of $x$, then the field equation
with the 4D $\Psi $ as AQ, (\ref{mecp}), becomes
\begin{equation}
v(\partial \cdot \Psi )-(v\cdot \partial )\Psi +i\left[ v\wedge (\partial
\wedge \Psi )\right] I=0.  \label{dc}
\end{equation}
Then we introduce a generalization of the correspondence principle that
deals with 4D AQs
\begin{equation}
i\text{%
h\hskip-.2em\llap{\protect\rule[1.1ex]{.325em}{.1ex}}\hskip.2em%
}\partial \rightarrow p.  \label{c1}
\end{equation}
Inserting (\ref{c1}) into (\ref{dc}) we reveal Dirac-like relativistic wave
equation for the free photon, which is written with AQs
\begin{equation}
v(p\cdot \Psi )-(v\cdot p)\Psi +i\left[ v\wedge (p\wedge \Psi )\right] I=0.
\label{Da}
\end{equation}
If we write Eq. (\ref{dc}) with CBGQs in the standard basis $\left\{ \gamma
_{\mu }\right\} $ then we get an equation that is very similar to (\ref{gako}%
)
\begin{equation}
\lbrack (\Gamma ^{\alpha })_{\ \mu }^{\beta }(\partial _{\alpha }\Psi ^{\mu
})]\gamma _{\beta }=0.  \label{g}
\end{equation}
Remember that $v$ in (\ref{dc}) and (\ref{g}) is independent of $x,$ whereas
$(\Gamma ^{\alpha })_{\ \mu }^{\beta }$ is the same as in (\ref{gako}). When
the generalized correspondence principle (\ref{c1}) is written with CBGQs in
the $\left\{ \gamma _{\mu }\right\} $ basis it takes the form
\begin{equation}
\gamma ^{\alpha }i\text{%
h\hskip-.2em\llap{\protect\rule[1.1ex]{.325em}{.1ex}}\hskip.2em%
}\partial _{\alpha }\rightarrow \gamma ^{\alpha }p_{\alpha }.  \label{co}
\end{equation}
Inserting (\ref{co}) into Eq. (\ref{g}) we find the following equation
\begin{equation}
\lbrack (\Gamma ^{\alpha })_{\ \mu }^{\beta }(p_{\alpha }\Psi ^{\mu
})]\gamma _{\beta }=0.  \label{dr}
\end{equation}
The equation (\ref{dr}) is Dirac-like relativistic wave equation for the
free photon, but now written with CBGQs in the $\left\{ \gamma _{\mu
}\right\} $ basis.

It is clear from the form of the equations (\ref{meko}), (\ref{gako})\ (with
some general $v^{\mu }$) and (\ref{g}), (\ref{dr}) (with $v^{\mu }$
independent of $x$) that they are invariant under the passive LT, since
every 4D CBGQ is invariant under the passive LT. The field equations with
primed quantities, thus in a relatively moving $S^{\prime }$ frame, are
exactly equal to the corresponding equations in $S,$ which are given by the
above mentioned relations. Thus these equations are not only covariant but
also the Lorentz invariant field equations. The principle of relativity is
automatically included in such formulation.

In addition let us briefly examine how one can get the field equations in
the formulation with 1-vectors of the electric field $E$ and the magnetic
field $B$ from the corresponding Eq (\ref{mecp}), when AQs are used, and (%
\ref{meko}) or (\ref{gako}), when CBGQs in the $\left\{ \gamma _{\mu
}\right\} $ basis are used. These equations are already obtained and
discussed in detail in [9] and previously in [7]. Substituting the
decomposition of $\Psi $ into $E$ and $B$ from (\ref{pkom}) into (\ref{mecp}%
) one gets two equations with real multivectors
\begin{eqnarray}
\partial \cdot (E\wedge v)+\left[ \partial \wedge (cB\wedge v)\right] I
&=&j/\varepsilon _{0}.  \notag \\
\partial \cdot (cB\wedge v)-\left[ \partial \wedge (E\wedge v)\right] I &=&0.
\label{re}
\end{eqnarray}
The equations (\ref{re}) are the same as Eq. (39) in [9]. Similarly,
starting with (\ref{meko}) we find
\begin{eqnarray}
\partial _{\alpha }(\delta _{\quad \mu \nu }^{\alpha \beta }E^{\mu }v^{\nu
}+\varepsilon ^{\alpha \beta \mu \nu }v_{\mu }cB_{\nu })\gamma _{\beta }
&=&(j^{\beta }/\varepsilon _{0})]\gamma _{\beta },  \notag \\
\partial _{\alpha }(\delta _{\quad \mu \nu }^{\alpha \beta }v^{\mu }cB^{\nu
}+\varepsilon ^{\alpha \beta \mu \nu }v_{\mu }E_{\nu })\gamma _{\beta } &=&0,
\label{re1}
\end{eqnarray}
which is the same as Eq. (40) in [9]. Of course, these equations, (\ref{re})
and (\ref{re1}), are also Lorentz invariant field equations but with
1-vectors $E$ and $B$. \medskip \bigskip

\noindent \textbf{5. Comparison with Majorana-Maxwell equations with }

\textbf{the 3D }$\mathbf{\Psi }$ \textbf{\bigskip }

\noindent Let us now see how our results can be reduced to Majorana-Maxwell
equations with the 3D $\mathbf{\Psi }.$ In the presence of sources these
equations are
\begin{equation}
div\mathbf{\Psi =}\rho /\varepsilon _{0},\quad irot\mathbf{\Psi }=\mathbf{j}%
/\varepsilon _{0}c+(1/c)\partial \mathbf{\Psi }/\partial t,  \label{ma}
\end{equation}
see, e.g., Eqs. (2) in [1]. When the sources are absent, $\rho =0$, $\mathbf{%
j}=\mathbf{0}$, and when the correspondence principle $W\rightarrow i$%
h\hskip-.2em\llap{\protect\rule[1.1ex]{.325em}{.1ex}}\hskip.2em%
$\partial /\partial t$, $\mathbf{p}\rightarrow -i$%
h\hskip-.2em\llap{\protect\rule[1.1ex]{.325em}{.1ex}}\hskip.2em%
$\mathbf{\nabla }$ is used in (\ref{ma}), then Eq. (\ref{ma}) with the 3D $%
\mathbf{\Psi }$ leads to the transversality condition and to
Majorana-Maxwell equation in a Dirac-like form
\begin{equation}
\mathbf{p}\cdot \mathbf{\Psi }=0,\quad W\mathbf{\Psi }+i\mathbf{p\times \Psi
}=\mathbf{0,}  \label{mm}
\end{equation}
see Eqs. (43) and (44) in [2].

As seen from (\ref{f}) (or (\ref{pkom}) and (\ref{itf})) the complex
1-vectors $\Psi $ and $\overline{\Psi }$ (or 1-vectors $E$ and $B$) are not
uniquely determined by $F$, but their explicit values depend also on $v$.
Let us choose the frame in which the observers who measure $\Psi $ and $%
\overline{\Psi }$, i.e., $E$ and $B$, are at rest. For them $v=c\gamma _{0}$%
. This frame will be called the frame of ``fiducial'' observers (for that
name see [16]), or the $\gamma _{0}$ - frame. All quantities in that frame
will be denoted by the subscript $f$, e.g., $\Psi _{f}$, $E_{f}$ and $B_{f}$%
. Furthermore, the standard basis $\left\{ \gamma _{\mu }\right\} $ will be
chosen in the $\gamma _{0}$ - frame. Then in that frame the velocity $%
v=c\gamma _{0}$ has the components $v^{\alpha }=(c,0,0,0)$ and $\Psi $ and $%
\overline{\Psi }$ ($E$ and $B$) become $\Psi _{f}$ and $\overline{\Psi }_{f}$
($E_{f}$ and $B_{f}$) and they do not have temporal components, $\Psi
_{f}^{0}=\overline{\Psi }_{f}^{0}=0$, $E_{f}^{0}=B_{f}^{0}=0$. In the $%
\gamma _{0}$ - frame Eq. (\ref{meko}) becomes
\begin{equation}
(\partial _{i}\Psi _{f}^{i}-j^{0}/\varepsilon _{0}c)\gamma
_{0}+(i\varepsilon _{\quad 0j}^{ki}\partial _{k}\Psi _{f}^{j}-\partial
_{0}\Psi _{f}^{i}-j^{i}/\varepsilon _{0}c)\gamma _{i}=0.  \label{mj}
\end{equation}
All terms in (\ref{mj}) are CBGQs that are written in the $\left\{ \gamma
_{\mu }\right\} $ basis. The equation (\ref{mj}) cannot be further
simplified as a geometric equation. However if one compares the components
from Eq. (\ref{mj}) and the components from Majorana-Maxwell equations (\ref
{ma}) then it is seen that they are the same. Hence it is the component form
of Eq. (\ref{mj}) (the ``fiducial'' frame and the standard basis $\left\{
\gamma _{\mu }\right\} $) which agrees with the component form of
Majorana-Maxwell equations with the 3D $\mathbf{\Psi }$ (\ref{ma}).

In the case when $j^{\mu }=0$, and with the replacement (\ref{co}), Eq. (\ref
{mj}) can be written as
\begin{equation}
p_{i}\Psi _{f}^{i}\gamma _{0}+(-p_{0}\Psi _{f}^{i}+i\varepsilon _{\quad
0j}^{ki}p_{k}\Psi _{f}^{j})\gamma _{i}=0.  \label{pe}
\end{equation}
The same result (\ref{pe}) follows from Eq. (\ref{dr}) when it is considered
in the $\gamma _{0}$ - frame in which the $\left\{ \gamma _{\mu }\right\} $
basis is chosen. The whole equation (\ref{pe}) is written with CBGQs in the
standard basis $\left\{ \gamma _{\mu }\right\} $ and cannot be further
simplified as a geometric equation. Comparing that equation with
Majorana-Maxwell equations (\ref{mm}) we again see that only component forms
of both equations can be compared. From the first term (with $\gamma _{0}$)
in Eq. (\ref{pe}) we find the component form of the transversality condition
written with 4D $p$ and $\Psi _{f}$, $p_{i}\Psi _{f}^{i}=0$ (remember that
in the $\gamma _{0}$ - frame $\Psi _{f}^{0}=0$), which agrees with the
component form of the transversality condition with the 3D $\mathbf{p}$ and $%
\mathbf{\Psi }$ from Eq. (\ref{mm}), $p_{i}\Psi ^{i}=0$. The second term
(with $\gamma _{i}$) in Eq. (\ref{pe}) yields the component form of
Dirac-like equation for the free photon that is written with 4D $p$ and $%
\Psi _{f}$. It agrees with the component form of the corresponding equation (%
\ref{mm}) with the 3D $\mathbf{p}$ and $\mathbf{\Psi }$.

Similarly, in the frame of ``fiducial'' observers and in the $\left\{ \gamma
_{\mu }\right\} $ basis, we can derive the component form of the usual
Maxwell equations with the 3D $\mathbf{E}$ and $\mathbf{B}$ from Eq. (\ref
{re1}). This is discussed in detail in [9].

However, it is worth noting that there are very important differences
between our Eqs. (\ref{mj}) and (\ref{pe}), or, better to say, our Eqs. (\ref
{g}) and (\ref{dr}), and Majorana-Maxwell equations (\ref{ma}) and (\ref{mm}%
). Our equations (\ref{mj}), (\ref{pe}), (\ref{g}) and (\ref{dr}) are
written with 4D CBGQs and the components are multiplied by the unit
1-vectors $\gamma _{\mu }$, whereas Majorana-Maxwell equations (\ref{ma})
and (\ref{mm}) are written with 3D vectors and the components are multiplied
by the unit 3D vectors $\mathbf{i}$, $\mathbf{j}$, $\mathbf{k}$. Only in the
frame of ``fiducial'' observers and in the $\left\{ \gamma _{\mu }\right\} $
basis the temporal component of the complex 1-vector $\Psi $ is zero, but in
all other relatively moving inertial frames this component is different from
zero. Furthermore in any frame other than the $\gamma _{0}$ - frame the
``fiducial'' observers are moving and the velocity $v$ has the spatial
components as well.

The complex 1-vector $\Psi $ transforms under the LT as every 1-vector
transforms, e.g., the components transform as in Eq. (\ref{ea}), whereas the
unit 1-vectors $\gamma _{\mu }$ transform by the inverse LT. This gives that
the whole $\Psi $ is unchanged, i.e., it holds that $\Psi ^{\mu }\gamma
_{\mu }=\Psi ^{\prime \mu }\gamma _{\mu }^{\prime }$ as for any other 4D
CBGQ. On the other hand there is no transformation which transforms the unit
3D vectors $\mathbf{i}$, $\mathbf{j}$, $\mathbf{k}$ into the unit 3D vectors
$\mathbf{i}^{\prime }$, $\mathbf{j}^{\prime }$, $\mathbf{k}^{\prime }$.
Hence it is not true that, e.g., the 3D vector $\mathbf{E}^{\prime }\mathbf{=%
}E_{1}^{\prime }\mathbf{i}^{\prime }+E_{2}^{\prime }\mathbf{j}^{\prime
}+E_{3}^{\prime }\mathbf{k}^{\prime }$ is obtained by the LT from the 3D
vector $\mathbf{E=}E_{1}\mathbf{i}+E_{2}\mathbf{j}+E_{3}\mathbf{k}$. Namely
the components $E_{i}$ of the 3D $\mathbf{E}$ are transformed by the usual
transformations (\ref{et}), which differ from the LT (\ref{ea}), and, as
said above, there is no transformation for the unit 3D vectors $\mathbf{i}$,
$\mathbf{j}$, $\mathbf{k}$. The same hold for the transformations of the 3D $%
\mathbf{B}$ and consequently for the transformations of the 3D $\mathbf{\Psi
}$. This means that the correspondence of the 4D picture with complex
1-vector $\Psi $ and the 3D picture with Majorana 3D complex vector $\mathbf{%
\Psi }$ exists only in the frame of ``fiducial'' observers and in the $%
\left\{ \gamma _{\mu }\right\} $ basis and not in any other relatively
moving inertial frame, or in some nonstandard basis. Moreover, that
correspondence in the $\gamma _{0}$ - frame and in the $\left\{ \gamma _{\mu
}\right\} $ basis refers only to the component forms of the corresponding
equations. Our equations with 4D geometric quantities are the same in all
relatively moving inertial frames, i.e., they are Lorentz invariant
equations, whereas it is not true for Majorana-Maxwell equations with the 3D
$\mathbf{\Psi }$.

Similarly it is proved in [9] that, contrary to the generally accepted
opinion, Maxwell equations with the 3D $\mathbf{E}$ and $\mathbf{B}$ are not
covariant under the LT. The field equations for the electric and magnetic
fields that are Lorentz invariant are, e.g., the equations with 1-vectors $E$
and $B$, Eqs. (\ref{re1}).

The situation with the physical importance of the 4D fields $\Psi $ and $%
\overline{\Psi }$ and the corresponding 3D fields $\mathbf{\Psi }$ and $%
\mathbf{\Psi }^{\ast }$ is the same as it is the situation with the physical
importance of the 4D fields $E$ and $B$ and the corresponding 3D fields $%
\mathbf{E}$ and $\mathbf{B}$. The comparison with experiments, the motional
electromotive force in [8], the Faraday disk in [9] and the Trouton-Noble
experiment in [10], strongly support our conclusions that the 4D fields $E$
and $B$ are not the ''auxiliary fields,'' as explicitly considered in [2]
and tacitly assumed in all previous works, but that an independent physical
reality must be attributed to such 4D fields $E$ and $B$ (or even better to
the electromagnetic field $F$, [10]) and not to the corresponding 3D fields $%
\mathbf{E}$ and $\mathbf{B}$. More generally, it is shown in [5] that there
is a true agreement, which is independent of the chosen reference frame and
the coordinate system in it, between the theory that deals with 4D geometric
quantities and the well-known experiments which test special relativity, the
''muon'' experiment, the Michelson-Morley type experiments, the
Kennedy-Thorndike type experiments and the Ives-Stilwell type experiments.
It is also discovered in [5] that, contrary to the common opinion, there is
no such agreement between Einstein's formulation of special relativity and
the mentioned experiments.\medskip \bigskip

\noindent \textbf{6. Conclusions \bigskip }

\noindent The consideration presented in this paper reveals that in the 4D
spacetime the complex fields, the 4D $\Psi $ and its complex reversion $%
\overline{\Psi }$, are physically important and well-defined quantities that
correctly transform under the LT, whereas it is not the case with the 3D
complex field $\mathbf{\Psi }$ and its complex conjugate field $\mathbf{\Psi
}^{\ast }$. In the 4D spacetime Majorana-Maxwell equations with the 3D $%
\mathbf{\Psi }$, (\ref{ma}) and (\ref{mm}), have to be replaced with our
Lorentz invariant field equations with the 4D $\Psi $.

For $j\neq 0$ we have presented new Lorentz invariant field equation (\ref
{mecp}) in which only the 4D AQs are used. Eqs. (\ref{meko}) and (\ref{gako}%
)\ are the corresponding field equations with 4D CBGQs written in the
standard basis $\left\{ \gamma _{\mu }\right\} $. For $j=0$ we have field
equations for the 4D $\Psi $, (\ref{dc}) and (\ref{Da}), with 4D AQs and (%
\ref{g}) and (\ref{dr}) with 4D CBGQs.

A new generalization of the correspondence principle is introduced by Eq. (%
\ref{c1}), where the AQs are used, or by Eq. (\ref{co}) with CBGQs.

The equations (\ref{Da}) (with AQs) and (\ref{dr}) (with CBGQs) are new
forms for Dirac-like relativistic wave equations for the free photon, which
are not yet reported in the literature, as I am aware. They will be the
starting point for the construction of the observer independent
stress-energy vector $T(n)$ (1-vector) and all other quantities that are
derived from $T(n)$, as are the energy density $U$ (scalar, i.e., grade-0
multivector), the Poynting vector $S$ (1-vector), etc. All these quantities
will be expressed by means of the 4D $\Psi $ and $\overline{\Psi }$ in a
complete analogy with the construction of these quantities in the axiomatic $%
F$ formulation [10].\bigskip

\noindent \textbf{References\bigskip }

\noindent \lbrack 1] R. Mignani, E. Recami and M. Baldo, ``About a
Dirac-like equation for the

photon according to Ettore Majorana'', Lett. Nuovo Cimento \textbf{11}, 568
(1974).

\noindent \lbrack 2] S. Esposito, ``Covariant Majorana formulation of
electrodynamics'', Found.

Phys. \textbf{28}, 231 (1998).

\noindent \lbrack 3] A. Einstein, ``On the electrodynamics of moving
bodies'', Ann. Phys. \textbf{17},

891 (1905), tr. by W. Perrett and G.B. Jeffery, in The Principle of
Relativity

(Dover, New York, 1952)..

\noindent \lbrack 4] T. Ivezi\'{c}, ````True transformations relativity''\
and electrodynamics'', Found.

Phys. \textbf{31}, 1139 (2001).

\noindent \lbrack 5] T. Ivezi\'{c}, ``An invariant formulation of special\
relativity, or the ''true

transformations relativity,'' and comparison with experiments'', Found. Phys.

Lett. \textbf{15}, 27 (2002); physics/0103026, (with 'radio' synchronization
as well).

\noindent \lbrack 6] T. Ivezi\'{c}, ``The proof that the standard
transformations of E and B are not

the Lorentz transformations'' Found. Phys. \textbf{33}, 1339 (2003).

\noindent \lbrack 7] T. Ivezi\'{c}, ``Invariant relativistic
electrodynamics. Clifford algebra approach'',

hep-th/0207250.

\noindent \lbrack 8] T. Ivezi\'{c}, ``The difference between the standard
and the Lorentz

transformations of the electric and magnetic fiels. Application to

motional EMF'', Found. Phys. Lett. \textbf{18}, 301 (2005).

\noindent \lbrack 9] T. Ivezi\'{c}, ``The proof that Maxwell's equations
with the 3D E and B are not

covariant upon the Lorentz transformations but upon the standard

transformations. The new Lorentz-invariant field equations'', Found.

Phys. \textbf{35,} 1585 (2005).

\noindent \lbrack 10] T. Ivezi\'{c}, ''Axiomatic geometric formulation of
electromagnetism with only

one axiom: the field equation for the bivector field $F$ with an explanation

of the Trouton-Noble experiment'' Found. Phys. Lett. \textbf{18}, 401 (2005).

\noindent \lbrack 11] D. Hestenes, Space-Time Algebra\textit{\ }(Gordon and
Breach, New York, 1966);

Space-Time Calculus\textit{; }available at: http://modelingnts.la.
asu.edu/evolution.

html; New Foundations for Classical Mechanics (Kluwer, Dordrecht,

1999) 2nd. edn.; Am. J Phys. \textbf{71}, 691 (2003).

\noindent \lbrack 12] C. Doran, and A. Lasenby, Geometric algebra for
physicists\textit{\ }(Cambridge

University, Cambridge, 2003).

\noindent \lbrack 13] D. Hestenes and G. Sobczyk, Clifford Algebra to
Geometric Calculus

(Reidel, Dordrecht, 1984).

\noindent \lbrack 14] J.D. Jackson, Classical Electrodynamics (Wiley, New
York, 1977) 2nd edn.

\noindent \lbrack 15] M. Riesz, Clifford Numbers and Spinors, Lecture Series
No. 38, The

Institute for Fluid Dynamics and Applied Mathematics, University of

Maryland (1958).

\noindent \lbrack 16] H.N. N\'{u}\~{n}ez Y\'{e}pez, A.L. Salas Brito, and
C.A. Vargas, ``Electric and

magnetic four-vectors in classical electrodynamics'', Revista Mexicana

de F\'{i}sica \textbf{34}, 636 (1988).

\end{document}